\documentclass[letterpaper]{article}
\usepackage{prepr}
\usepackage{amsmath,amsfonts} %%assymb breaks the acm template
\usepackage{url}
\usepackage{booktabs} % For formal tables                  
\usepackage{graphicx}
\usepackage{textcomp}
\usepackage{xcolor}
\usepackage{xspace}
\usepackage{hyperref}
\usepackage{balance}
\usepackage{colortbl}
\usepackage{tabularx}
\usepackage{tabulary}
\usepackage{booktabs}
\usepackage{multicol}
\usepackage{multirow}
\usepackage{rotating}
\usepackage{caption}
\usepackage{subcaption}
\usepackage[export]{adjustbox}
\usepackage{tabularx}
\usepackage{tikz}
\usepackage{float}
\usepackage[shortlabels]{enumitem}
\usepackage[most]{tcolorbox}
\usepackage{xparse}
\usepackage{etoolbox}
\usepackage{xurl}

\definecolor{commentGreen}{RGB}{128,127,41}
\definecolor{stringGreen}{RGB}{0,127,38}
\definecolor{keywordRed}{RGB}{151,0,11}
\definecolor{typePurple}{RGB}{128,3,82}
\definecolor{variableBlue}{RGB}{0,0,255}
\definecolor{annotationRed}{RGB}{198,4,38}
\definecolor{annotationBlue}{RGB}{32,21,223}

\definecolor{01gen}{RGB}{245,245,245}
\definecolor{02los}{RGB}{213,232,212}
\definecolor{03act}{RGB}{225,213,231}
\definecolor{04ref}{RGB}{218,232,252}
\definecolor{05misc}{RGB}{255,242,204}

\newenvironment{keywords}[1]
{
  \par\noindent\textbf{Keywords: }%
  \@keywords#1,\@nil
  \par
}{}

\def\@keywords#1,#2\@nil{%
  #1%
  \if\relax\detokenize{#2}\relax
  \else
    ~\textbullet~%
    \@keywords#2\@nil
  \fi
}

\newtcolorbox[auto counter]{reqbox}[1]{title={\bfseries #1},
  coltitle=white, 
  leftrule=0.25mm,
  rightrule=0.25mm,
  bottomrule=0.25mm,
  toprule=0.25mm,
  colframe=white!45!black, boxsep=2pt,left=4pt,right=4pt,top=3pt,bottom=3pt,  sharpish corners}
\usepackage{tabularx}
\usepackage{booktabs}
\title{Design and Deployment of a Course-Aware AI Tutor in an\\ Introductory Programming Course}

\author{
\PREPauthor{Iris Groher}{Johannes Kepler University Linz, Institute of Business Informatics -- Software Engineering}{iris.groher@jku.at}
\PREPauthor{Patrick Heissenberger}{Johannes Kepler University Linz,  Institute of Business Informatics -- Software Engineering}{k12105284@students.jku.at}
\PREPauthor{Michael Vierhauser}{University of Innsbruck, Department of Computer Science}{Michael.Vierhauser@uibk.ac.at}
}

\setvenue{\textbf{18th International Conference on Computer Supported Education}}
\setvenueone{the \textbf{18th Int'l Conf. on Computer Supported Education}}
\setyear{2026}
\setdoi{}

\begin{document}
\pagestyle{custom}
\maketitle

\begin{abstract}
Large Language Models (LLMs) have become part of how students solve programming tasks, offering immediate explanations and even full solutions. Previous work has highlighted that novice programmers often  heavily rely on LLMs, thereby neglecting their own problem-solving skills. To address this challenge, we designed a course-specific online Python tutor that provides retrieval-augmented, course-aligned guidance without generating complete solutions. The tutor integrates a web-based programming environment with a conversational agent that offers hints, Socratic questions, and explanations grounded in course materials.
Students used the system during self-study to work on homework assignments, and the tutor also supported questions about the broader course material. We collected structured student feedback and analyzed interaction logs to investigate how they engaged with the tutor’s guidance. We observed that students used the tutor primarily for conceptual understanding, implementation guidance, and debugging, and perceived it as a course-aligned, context-aware learning support that encourages engagement rather than direct solution copying.
\end{abstract}


\section{Introduction}
\label{sec:intro}

Teaching introductory programming remains one of the ongoing challenges in computer science education~\cite{medeiros2018systematic}. This involves, for example, dealing with large cohorts and heterogeneous prior knowledge of students~\cite{Groher2022,Frankford2025}. This issue is further exacerbated by the inherently abstract nature of concepts taught in these courses, with previous work showing that students frequently struggle in translating natural language task descriptions into working code, beyond just conceptual problem comprehension~\cite{Lister2004}. These challenges are especially evident in teaching contexts with limited opportunities for individual guidance (e.g., due to large student numbers), particularly when students depend heavily on self-study~\cite{Denny2024}.

The emergence of Large Language Models (LLMs) has significantly transformed the way how students engage with programming tasks~\cite{Lyu2024}. Tools such as ChatGPT or GitHub CoPilot now provide code explanations, examples, and even full solutions, without the need to actually understand the conceptual elements behind it~\cite{Groothuijsen2024}. 
 Although these systems lower access barriers, they introduce a number of new problems. Existing work highlights the problems that emerge when students overly rely on LLM-generated code without developing their own coding skills, such as lack of engagement in problem-solving processes or lack of conceptual knowledge in the context of programming~\cite{Lehmann2024}. 

In response, LLM-driven tutoring systems have emerged that build guardrails around the information provided to students. Rather than providing complete solutions, these systems guide students through structured hints and stepwise problem-solving support~\cite{Roest2024}. Recent work confirms the potential of such customized tutors that are explicitly aligned with course contents and materials, which can adapt explanations to students’ competence levels, offer context-sensitive guidance, and scale to large cohorts~\cite{Li2025,Teng2025}. However, empirical evidence on their classroom integration remains limited, particularly regarding concrete design decisions, implementation practices, and observed patterns of student interaction.

This paper presents the design and deployment of a course-specific online Python tutor that integrates a web-based programming environment, retrieval-augmented access to the course’s teaching materials, and a conversational agent. The system is designed to (1) provide feedback grounded exclusively in the official course materials, (2) encourage algorithmic thinking through hints and Socratic questions rather than direct code generation, and (3) support students during self-study at home by helping them work on their homework assignments.
 It was deployed in an introductory Python course and integrated into self-study tasks on collections and functions—topics identified as particularly challenging—while remaining applicable across the full course scope. To evaluate its use, we collected structured student feedback and analyzed interaction logs to examine how students engaged with the tutor’s guidance.

This paper contributes \emph{(i) the design and implementation of a course-aware LLM-based programming tutor} aligned with course objectives and materials, \emph{(ii) an empirical evaluation in a classroom setting}, combining interaction log analysis and survey data, and \emph{(iii) lessons learned} from its development, deployment, and use in practice.

\vspace{1em}\textbf{Data Availability:}
{\small\url{https://github.com/TeachingAndLearningSciences/PythonTutor}}
%\vspace{-0.5em}

\begin{figure*}[t!]
  \centering
  \includegraphics[width=0.95\textwidth]{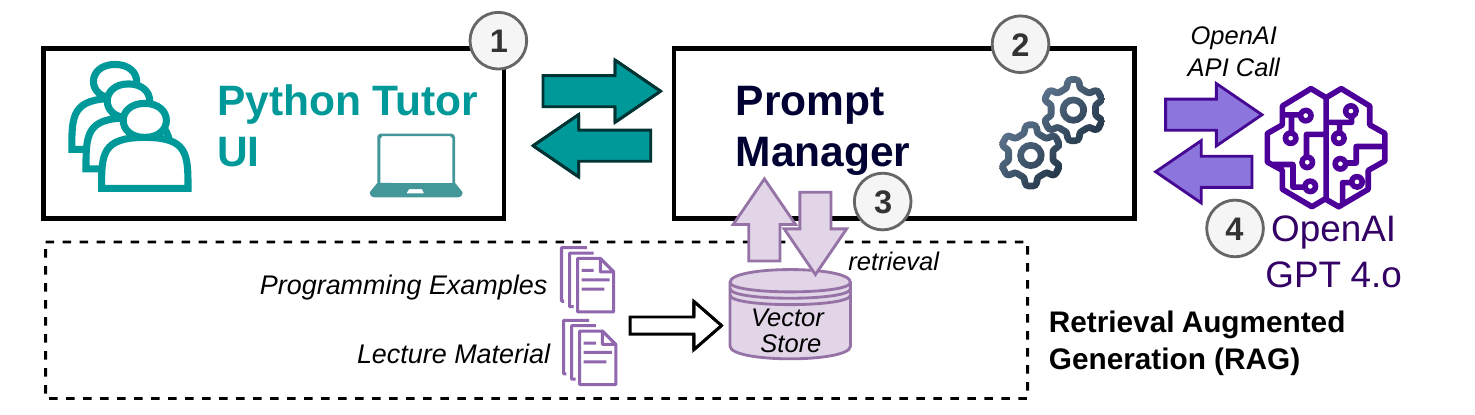} %\vspace{-1em}
  \caption{Architecture of the Python Online Tutor}
  \label{fig:tutor}
  %\vspace{-1.8em}
\end{figure*}

\section{Background \& Related Work}
\label{sec:background}

Studies have shown that AI coding assistants, when embedded in instructional tools, can effectively support learning processes, support broader engagement in programming~\cite{Kazemitabaar2023}, and significantly improve scores~\cite{Lyu2024}.

Before the emergence of LLMs, educational chatbots ranged from simple keyword-based systems to more sophisticated rule-driven agents. Early systems such as ELIZA~\cite{Weizenbaum1966} demonstrated the potential of conversational interaction, while later Intelligent Tutoring Systems introduced personalized feedback, but remained limited by narrow domain models and high development costs~\cite{Crow2018}.
Beyond this, LLMs now provide new opportunities for flexible, conversational support in programming education. First studies and analyses show that general-purpose code generation tools, including GPT-based models, can, on the one hand, support students by, for example,  generating explanations or example solutions, but, on the other hand, also introduce risks such as incorrect code, misalignment of LLM feedback with the current learning contents, and over-reliance~\cite{Becker2023,Hellas2023}. These concerns have also triggered new research in pedagogically constrained, context-aware, and solution-avoiding tutoring systems. 

Several studies focus on domain-specific, pedagogically aligned agents. For example, \textit{CodeHelp}~\cite{Liffiton2024}, a web-based coding assistant,  generates tailored hints for code and related questions provided by students. The tutor guides and explains, but does not provide solution code. Similarly, Kazemitabaar et al.~\cite{Kazemitabaar2024} developed \textit{CodeAid}, an LLM-based assistant that provides conceptual explanations, pseudo-code, and annotated feedback on student errors, while also avoiding complete solutions. Tools such as the \textit{CS50 Duck} encourage self-guided learning but rely on students to provide exercise descriptions and code snippets to get relevant support~\cite{Liu2024}. Bassner et al.~\cite{Bassner2024} present \textit{Iris}, a chat-based virtual tutor integrated into the Artemis learning platform~\cite{krusche2018artemis} that provides personalized, context-aware support for programming exercises. Iris is designed to align with course pedagogy by avoiding full solutions and instead offering hints and counter-questions that promote independent problem solving. Lyu et al.~\cite{Lyu2024} conducted a study with 50 students to examine how an LLM-based assistant (\textit{CodeTutor}) influences learning outcomes, attitudes, and help-seeking behaviors in an introductory programming course. Students using CodeTutor achieved significantly higher final scores but also expressed concerns about its limited support for critical thinking.
Although existing systems focus primarily on assistance during interactive programming exercises, little is known about how LLM-based tutors can also support self-study, a context in which students often work without access to timely human feedback. Moreover, although several systems emphasize guardrails, fewer combine retrieval-augmented access to course-specific materials with Socratic prompting to reinforce conceptual reasoning. Recent work also shows that when students use LLM-based assistants in programming courses, most queries focus on debugging and implementation rather than conceptual understanding~\cite{Sheese2024}, highlighting the need for tutors that foster deeper reasoning and align feedback with course learning objectives.

Our work contributes to this by introducing a course-aligned Python tutor designed to support students both during self-study and while working on their programming assignments. The provided support remains fully consistent with the course-defined learning objectives.

\section{Python Online Tutor}
\label{sec:approach}

In the following, we present our Python Online Tutoring framework. We first provide a brief overview and then describe the core components in detail.

\subsection{Requirements}
\label{sec:req}
Based on existing literature~\cite{Denny2024}, and our own experience in programming education, we derived five core requirements (REQ1-REQ5) for the tutoring system, and additional non-functional requirements.

\bullitem{REQ1 -- Context-Aware Interaction}~\cite{shum2023personalised,barbosa2024adaptive}:
\emph{The agent shall be integrated directly into the programming environment to ensure continuous access to the actual working context.}
It must be capable of capturing and processing the currently visible or edited source code, as well as relevant task descriptions provided within the development environment. %Code changes shall be detected in real time to ensure that the agent always operates on the current state.

\bullitem{REQ2 -- Personalized Conversational Behavior}~\cite{lim2025learning,farhood2025artificial}:
\emph{The agent shall adapt its responses to individual users and ongoing conversations.}
It must support follow-up interactions and account for prior dialogue history. Based on prior knowledge, the agent shall also dynamically adjust the level of detail to avoid overly simplistic or excessively complex explanations.

\bullitem{REQ3 -- Course-Conform Responses}~\cite{Liffiton2024,dong2025build}:
\emph{The agent shall generate responses exclusively based on the materials used in the course.} Only concepts, methods, and terminology explicitly covered in the course may be explained or referenced. Queries related to specific lectures or exercises must be answered using the corresponding course materials. Requests outside the defined course scope shall be explicitly rejected with a notification.

\bullitem{REQ4 -- Didactic Guidance}~\cite{shum2023personalised}:
\emph{The agent shall support learning through didactically appropriate behavior.} Instead of providing complete solutions or a full implementation, it must offer hints, guiding questions, and conceptual prompts tailored to the specific problem. The depth and number of hints shall be dynamically adjusted based on the dialogue history and 
inferred knowledge.

\bullitem{REQ5 -- Configurability}~\cite{anaroua2025ai,farhood2025artificial}:
\emph{The agent shall support runtime configuration, particularly with respect to different levels of context awareness.} This, particularly pertains to code and task awareness, meaning that these changes must take effect immediately without requiring a system restart.

Besides the core functionality,  we considered \emph{Usability}, 
as the tutor needs to be intuitive and easy to use, also for users without advanced technical expertise; \emph{Performance}, to provide responses quickly and reliably to user interactions, and scalable and efficient resource management to handle responses under parallel load. 
Additionally, as with any AI-based system using LLMs, ethical considerations need to be taken into consideration. This aspect is particularly important when students are working with these systems, ensuring that no discriminatory, offensive, or harmful content is generated.

Finally, to support validation (cf.~\citesec{validation}), we explicitly included \emph{Monitoring and Logging}  to log token usage, associated costs, prompts, thread identifiers, and active context-awareness settings. This includes automated data collection that meets privacy requirements.

\subsection{Architecture Overview}

In the following, we provide a brief description of our architecture (cf.~\citefig{tutor}), the prototype implementation, and how the previously described requirements were incorporated in our tutoring system.

Students access the system directly through a \emph{Web-based frontend (1)} using their personal devices. The frontend integrates the AI tutor, a simple programming environment, as well as task descriptions within a single user interface. When a student interacts with the tutor, the entered prompt and, depending on the configured level of context awareness, the task description together with the currently edited source code, are transmitted to the backend \emph{Prompt Manager (2)}. 
As part of this, each request is assigned a unique thread identifier, enabling user-specific session management and coherent handling of follow-up prompts within the same conversational context.

The LLM generates a response grounded in the course materials, \emph{using Retrieval-Augmented Generation (RAG) (3)} guided by predefined system prompts and fixed model parameters. The generated output is returned to the backend, which forwards the response to the corresponding frontend client. Subsequent prompts reuse the same thread identifier, ensuring consistent contextual interpretation across interactions.
To support multiple concurrent users, all backend processing is performed asynchronously. Depending on the selected context-awareness level, the backend applies prompt engineering techniques to enrich or restructure the input before forwarding it to the \emph{OpenAI API (4)}.

The architecture consists of three main parts: the frontend, the Prompt Manager, and the LLM, accessed via the API.

\bullitem{Frontend:}
The frontend serves as the primary interface between students and the system and is executed client-side within the web browser. It manages sessions and stores essential parameters such as the thread identifier to associate user interactions with an active session.  In addition, the frontend allows users to configure context-related parameters, including the context-awareness level, which determines whether task descriptions and source code are included in requests. Students can submit prompts and receive responses directly within the integrated code editor. The frontend also displays task descriptions and presents system feedback such as status messages or error notifications.

\bullitem{Prompt Manager:} The backend represents the central control and processing unit of the system. It receives prompts, context settings, task descriptions, and source code from the frontend, manages session identifiers, and orchestrates the interaction with the OpenAI API.
Key responsibilities include asynchronous request handling, prompt optimization based on the selected context-awareness level, error handling, and response routing. The backend ensures scalability and robustness under concurrent usage while maintaining consistent conversational contexts across follow-up interactions.

\bullitem{OpenAI API (LLM):} The OpenAI endpoint provides the response generation. The backend submits preprocessed and contextualized prompts to the API. Course materials are embedded via retrieval-augmented generation to ensure that responses are grounded exclusively in authorized instructional content. System prompts and model parameters (e.g., temperature and maximum token limits) are predefined at the API level to ensure consistent response behavior. The generated responses are returned to the backend and subsequently delivered to the frontend.

\subsection{Prototype Implementation}
The frontend is implemented using React (cf.~\citefig{screenshot}), facilitating the development of interactive user interfaces, which is essential for concurrently managing the chat agent, embedded programming environment, and configuration controls.
The backend is implemented in Python due to its widespread adoption in education and research, as well as its extensive ecosystem of libraries. The FastAPI framework is used to implement a RESTful API with native support for asynchronous request handling, enabling efficient processing of concurrent interactions.
 The OpenAI Python library is used for integrating the LLM, using the OpenAI Assistant API in combination with an attached vector store, following the RAG paradigm. This enables the integration of domain-specific course materials into the response generation process without fine-tuning the model. Semantic retrieval ensures that responses are derived exclusively from authorized course documents, maintaining curricular alignment.

The tutor’s retrieval component is based on a knowledge base that contains the complete course materials, including lecture slides, annotated code examples, assignment descriptions, and explanatory text used in the course.
Although the evaluation described in this paper focuses on tasks related to collections and functions, the underlying RAG is not restricted to these topics and can provide guidance throughout the course.
%\vspace{-0.2em}

\section{Application Study}
\label{sec:validation}

In the following, we present the research method and research questions guiding our application study and describe the introductory Python course setting in which the tutor was deployed.

\begin{figure}[b!]
  \centering
  \includegraphics[width=1.0\columnwidth]{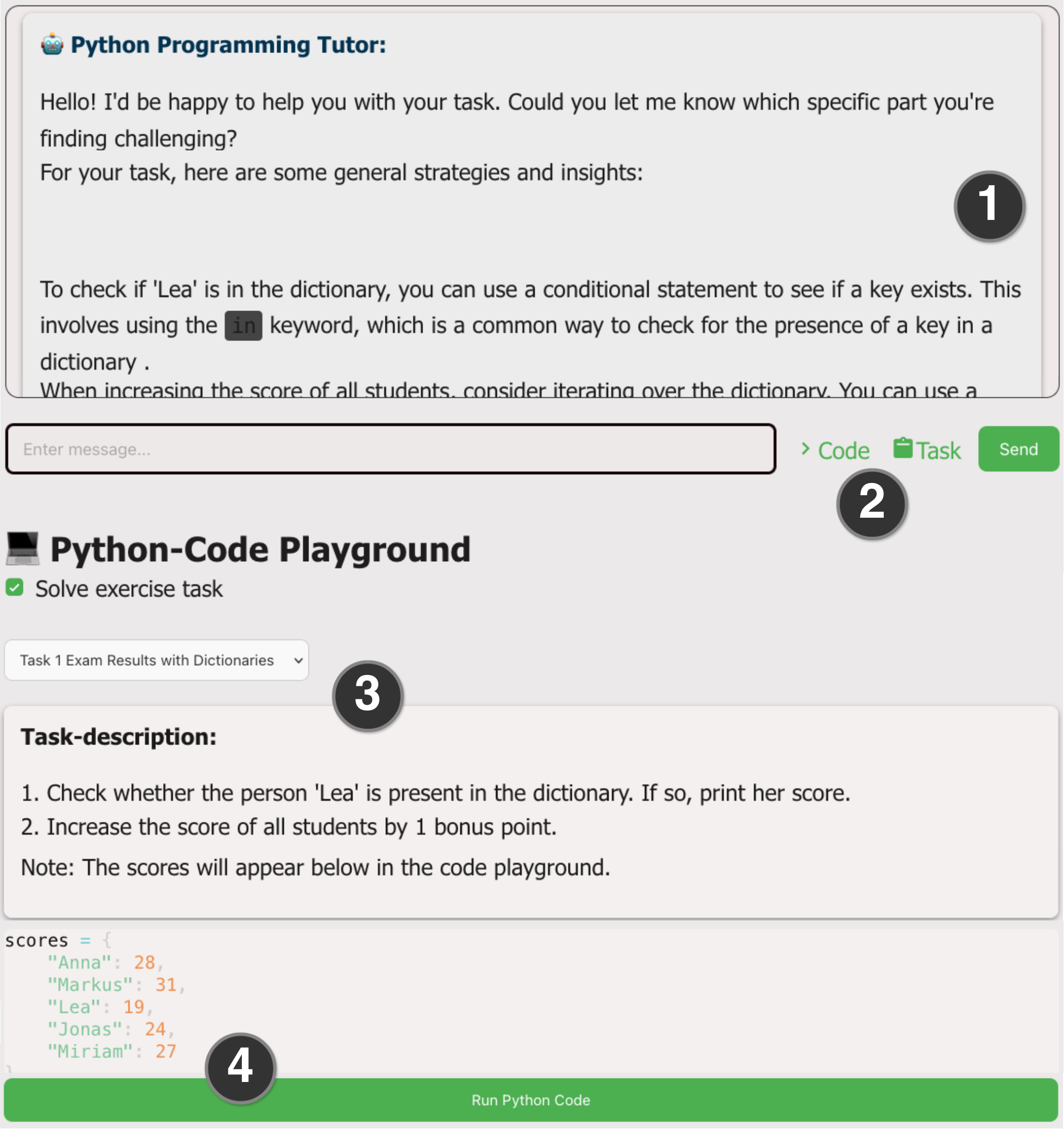}
  \caption{Tutor Frontend (1 = Prompt and Answer, 2 = Context Awareness, 3 = Task Description, 4 = Current Solution)}
  \label{fig:screenshot}
  %\vspace{-1.7em}
\end{figure}

\subsection{Research Method and Questions}

The tutor was deployed in a first-year,  introductory Python course for business administration students at our university.

%Further detail are provided in~\citesec{coverview}. 
The study follows a mixed-methods exploratory field study design~\cite{storey2024guiding}, combining interaction log analysis with survey data to investigate how the tutor is used and perceived by students. 
Students were introduced to the tutor through (1) a short instructional video explaining its purpose and use, and (2) a brief in-class presentation by the course instructor. As part of the study, students were asked to solve two exercise tasks, one on collections and one on functions, using the tutor. These were selected as previous iterations of the course have shown that students frequently struggle with these topics.
Limiting the evaluation to this subset of topics allowed us to study student interactions with the tutor in a controlled setting while keeping the assignments comparable across participants. However, students were free to ask questions about any of the other topics covered in the course.
The tutor was evaluated as a supporting tool, where students were independently working on homework programming assignments. In addition, students had access to the tutor for a period of four weeks, during which they could use it voluntarily for any course-related questions, including clarification of concepts or debugging support. No restrictions were imposed on the number or types of questions students could ask. All interactions were logged, including questions asked, as well as metadata (timestamps and task context).

After this period, students were asked to complete an online survey containing a total of 33 questions including demographic and background items (gender, prior programming experience, and general motivation), followed by a set of Likert-scale questions addressing the tutor’s didactic behavior (e.g., balance between hints and solutions, appropriateness to prior knowledge), context awareness (consideration of conversation history, task description, and their combination), and course alignment (consistency with course content and perceived reliance on course materials). 
Additional items focused on usability and interaction quality, including ease of use, clarity of language, trust in the answers, and technical reliability. Finally, an open-ended question asked students to elaborate on positive aspects and perceived shortcomings. The full questionnaire can be found in our online supplemental material.

Based on the study design and available data, we answer the following research questions:

\begin{itemize}[leftmargin=1.5em]
      \item \textbf{RQ1.} What types of questions do students ask when using an LLM-based tutor during independent work in an introductory Python course?
    \item \textbf{RQ2.} How do students perceive the usefulness and learning support provided by the tutor when solving programming tasks for challenging topics such as collections and functions?
\end{itemize}

To answer RQ1, student–tutor interaction logs were analyzed using an open coding approach to categorize the types of questions students asked. Only questions submitted without activated task or code context were considered. 
The coding scheme was developed iteratively through multiple rounds of analysis. Three researchers independently coded the students’ questions to the tutor in an initial round. Differences in coding were then systematically discussed and resolved until a stable set of categories was established.
To answer RQ2, the survey data was analyzed descriptively. Likert-scale items were summarized, and open-ended survey responses were analyzed thematically.

\subsection{Course Overview and Study Setup}
\label{sec:coverview}
The introductory Python module consists of a 3 ECTS lecture and a 3 ECTS exercise class, corresponding to a combined workload of 150 hours. 
The lecture introduces fundamental programming concepts (conditions, loops, collections, functions, modules) and provides self-study materials (slides, videos, code examples), as well as in-person recap meetings that focus on discussion, clarification of questions, and examples. The exercise complements the lecture through weekly programming practice. 
%Students complete eight graded assignments, each consisting of five tasks using Google Colab for coding with automatic evaluation in Moodle. 
Students complete eight graded assignments, each consisting of five tasks using Google Colab for coding with automatic evaluation in Moodle. Moodle is used as the central learning management system to distribute assignments and provide automated feedback on correctness via predefined test cases. The AI tutor operates independently of the grading system and serves as an optional support tool during self-study.
The exercise sessions include discussion of lecture content and walkthroughs of assignment solutions. Assignments cover core introductory topics such as variables, control flow, loops, functions, lists, and dictionaries, with later tasks increasing in complexity.

The study involved a total of 39 participants, with 13 participants identifying as male, 25 as female, and one participant not disclosing their gender. Regarding prior experience with programming, the majority of participants (26 students) reported no prior programming experience, while 13 participants indicated that they had some form of previous exposure to programming. The study sample thus predominantly consisted of novice learners. Most participants (30 students) were attending the course for the first time, whereas 9 participants reported that they had attended the course before. Participants were also asked to self-assess their general motivation to learn programming. The majority reported a normal (18) to high (9) or very high (6) level of motivation. A smaller number indicated low (4) motivation, and 2 participants reported not being motivated at all. At the time of the survey, participants were working on one of two assignment topics. 22 students were working on tasks related to collections, while 17 students were working on functions.

\section{Results}
\label{sec:results}

In the following, we provide a comprehensive analysis of the analyzed usage data (RQ1) from our study and the perceived usefulness from the survey (RQ2).

%\subsection{Usage of the Tutor and Questions Asked (RQ1)}
\subsection{Usage of the Tutor (RQ1)}

RQ1 analyzes the types of questions students asked when interacting with the tutor in our introductory Python course.
Before coding, we merged the interactions that belonged together (i.e., where students, for example, only sent the first half of a question, immediately followed by the second part, or additional parts). This resulted in a set of 304 interactions for coding (from a total number of 354 interactions). After applying two rounds of coding and discussion, we identified  12  question categories (cf. Table~\ref{tab:categories}).

\begin{table}[h]
\centering
\caption{Category Overview}
\footnotesize
\renewcommand{\arraystretch}{1.01}
\label{tab:categories}
\begin{tabular}{L{2.5cm}L{4.9cm}R{0.4cm}}
\toprule
\textbf{Category} & \textbf{Description} & \textbf{\#} \\
\midrule
Give Example            & Asking for illustrative examples or demonstrations. & 13 \\\midrule
Follow Up              & Simple follow-up questions building on a previous response. (e.g., \emph{"Yes please"}) & 17 \\\midrule
Course Material        & Questions related to provided or referenced course content. & 5  \\\midrule
Code Correctness & Identifying, explaining, or fixing errors in code. & 34 \\\midrule
Explain Code    & Requests to explain code snippets or language syntax. & 13 \\\midrule
Explain Concept        & Conceptual explanations without direct relation to the task. (e.g., \emph{"How do functions work in Python?"}) & 61 \\\midrule
Explain Tasks /Detail   & Clarification of tasks or steps. & 19 \\\midrule
How To                 & Question specifically ``how'' something is done/implemented. & 38 \\\midrule
Implement              & Explicit request to implement a solution or functionality. & 47 \\\midrule
Unrelated              & Queries not related to the task/assignment. (e.g., \emph{"What is the weather like today?"}) & 23 \\\midrule
 Code Only               & Code without explanation/question. & 31 \\\midrule
Misc            & Other Queries. &3\\
\midrule
\textbf{Total} & & \textbf{304} \\
\bottomrule
\end{tabular}
\vspace{-10pt}
\end{table}

Overall, students primarily used the tutor to support conceptual understanding and problem-solving. The most frequent category was \textit{Explain Concept} (61 instances), indicating that students often asked for clarification of fundamental programming concepts (e.g., \emph{"How do for-loops work in Python?"}). 
Additionally, students also asked more targeted questions, how to solve a specific problem (\textit{How To}, 38), or support for code comprehension and \textit{code correctness and errors} (34), or requested explanations of existing code (\textit{Explain Code}, 13).
This confirmed that the tutor was, in fact, used as intended, as an on-demand assistant during coding. 
Additionally, requests for \textit{task clarification} (19) and \textit{examples} (13) further emphasized that the students used the tutor to better understand assignment requirements. Interaction patterns such as short \textit{follow-up questions} (17) and \textit{code-only} submissions (31) show repeated, context-dependent dialogues, where students often provided minimal prompts and built on previous context. 
However, even though the tutor was not supposed to be used for solution generation, many interactions included explicit requests for providing implementation (\textit{Implement}, 47). Since the final prompt was created after the students submitted their question, we explicitly tried to prevent the LLM from returning the full solution (cf.~\citesec{req}).
Additionally, we also observed \textit{Unrelated} queries (23) that were completely out of scope and 3 other queries that did not fit in any of the above categories (\textit{Misc}).

\begin{reqbox}{\small RQ1: Types of Questions Asked }{}
\small
Analyzing the logs and prompts revealed that students predominantly used the tutor for conceptual explanations, implementation guidance, and debugging support, suggesting that the tutor was properly used as actual learning support rather than as a simple solution generator.
\end{reqbox}

\subsection{Perceived Usefulness (RQ2)}

RQ2 investigates how students perceive the usefulness and learning support provided by the tutor when solving programming tasks on challenging introductory topics, specifically collections and functions. \citefig{eval} presents the mean values of all survey items, grouped by category, while Table \ref{tab:survey} summarizes the corresponding category-level means and the contents of the survey items. Overall, students perceived our Python tutor very positively, with a total mean of 3.84 across all survey items.

\begin{figure}[bh!]
  \centering
  \fbox{\includegraphics[width=0.9\columnwidth]{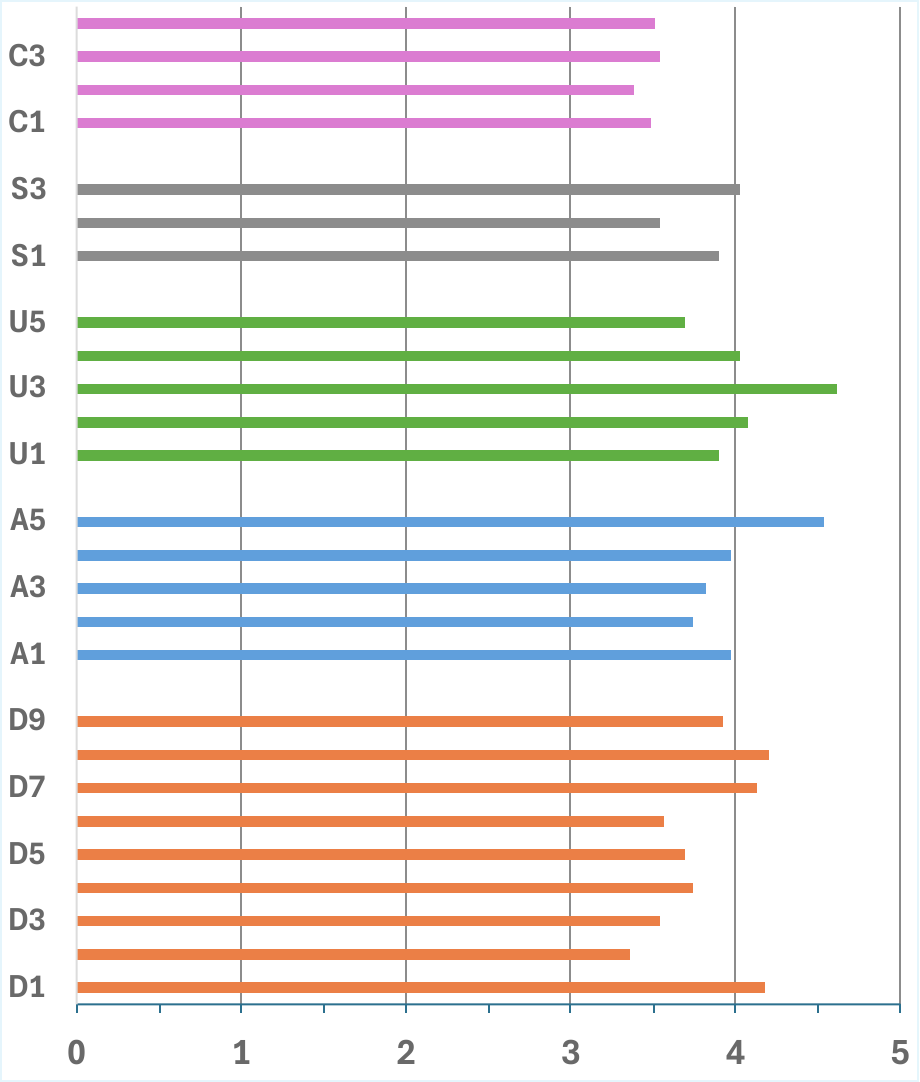}}
  \caption{Mean distribution of survey questions (orange = Didactics (D1-D9), blue = Awareness (A1-A5), green = Usability (U1-U5), grey = Summary (S1-S3), purple = ChatGPT (C1-C4))}
  \label{fig:eval}
\end{figure}

The \textbf{Didactics} category achieved a mean score of 3,81, reflecting generally positive perceptions of the tutor’s learning-oriented behavior. Students largely agreed that the chatbot avoided providing complete solutions and instead focused on hints and ideas (D1, M = 4,18), which aligns well with the goal of fostering independent problem solving. Students reported that the tutor’s consideration of prior conversation context (D7, M = 4,13) and its consistency with course content (D8, M = 4,21) were especially helpful.

Lower, but still positive, mean values were observed for items related to generating new solutions based on hints (D2, M = 3,36) and the perceived helpfulness of follow-up questions (D3, M = 3,54). This suggests that while many students benefited from the guidance, translating the provided hints into an actual solution remained challenging for some.

The \textbf{Awareness} category received a comparatively high mean score of 4,01, indicating that students clearly perceived the tutor’s ability to understand and integrate contextual information. Students reported that the tutor generally understood the current state of their code (A1, M = 3,97) and the task description (A3, M = 3,82), and that it could relate both aspects effectively (A4, M = 3,97).
The highest result was observed for A5 (M = 4,54), where students agreed that the combination of task and code (realized within the awareness selection options) would be helpful for future exercises. This highlights the perceived added value of a context-aware tutor.

\textbf{Usability} achieved the highest overall mean (4,06), indicating that interaction with the tutor was perceived as clear, and respectful. Students found the chatbot easy to use (U1, M = 3,90) and its language clear and understandable (U2, M = 4,08). In particular, the item addressing respectful and serious treatment of student input (U3) reached a very high mean (M = 4,62). Trust in the chatbot’s answers was also rated positively (U4, M = 4,03). Technical issues were reported only to a limited extent (U5, M = 3,69). 

\begin{table}[th!]
\centering
\caption{Survey Results from our feedback questionnaire.}
\small
\label{tab:survey}
\begin{tabular}{L{7.2cm}R{0.7cm}}  
\toprule
\textbf{Category}  & \textbf{Mean} \\
\midrule
\textbf{Didactics (D):}          
Whether independent learning through hints and follow-up questions was achieved, and  explanations were grounded in the course materials & 3,81 \\\midrule
\textbf{Awareness (A):}
Perceived effectiveness of the contextual awareness, and usefulness for future learning activities &  4,01 \\\midrule
\textbf{Usability (U):}
Ease of use, clarity, perceived trustworthiness, and technical issues  & 4,06 \\\midrule
\textbf{Summary (S):}  
Overall evaluation of working with the chatbot, impact on motivation, engagement, and learning & 3,82 \\\midrule
\textbf{ChatGPT (C):} 
Prior use of ChatGPT or similar LLMs, comparison of learning support and  preferences for solution generation versus guided learning & 3,48 \\
\midrule
\textbf{Total} &  \textbf{3,84} \\
\bottomrule
\end{tabular}
\end{table}

For the \textbf{Summary} category (M = 3,82), students also expressed a positive overall experience when working with the tutor (S1, M = 3,90), confirming that it encouraged more active thinking and experimentation (S3, M = 4,03). The perceived impact on motivation was rated lower but still positive (S2, M = 3,54).

Finally, for the \textbf{ChatGPT} category, results showed the lowest mean score (M = 3,48). Most participants had prior experience with ChatGPT or similar systems. When directly comparing the course-specific tutor to generic LLMs, students slightly favored the tutor in terms of learning support (C2, M = 3,38) and exam preparation or long-term understanding (C3, M = 3,54). %Responses to C4 (M = 3.51) indicate that students did not clearly prefer using ChatGPT for solution generation over working with the tutor.

As part of the last part, the free-text responses, several students explicitly highlighted the value of the explicit course alignment and code awareness of the tutor.
For example, one participant mentioned that \textit{"parts of my own code were taken into account -- that was very helpful"} and that they \textit{"particularly liked the prior knowledge of the course the bot has"}. Students also emphasized the tutor’s didactic quality and explanatory style, describing the interaction as \textit{“almost like a teacher guiding me step by step, without making the explanations too long or complicated.”} At the same time, the qualitative feedback reveals a current limitation related to system responsiveness. While the overall evaluation was clearly positive, multiple students mentioned that \textit{"the responses were sometimes a bit slow"}, which sometimes \textit{"interrupted the flow."}

\begin{reqbox}{\small RQ2: Perceived Usefulness }{}
\small
Our study results suggest that students generally perceived the tutor as supportive for learning, especially with respect to contextual awareness and course-aligned guidance, while fostering engagement without primarily encouraging direct solution copying.
\end{reqbox}

\subsection{Threats to Validity}
As with any study, our work is subject to several threats to validity.
 Concerning internal and construct validity, we had a limited number of 39 business administration students, with the majority having no prior programming experience. While this reflects a common target group for introductory programming courses, the results may not directly generalize to more advanced or technically experienced cohorts. Students with stronger prior knowledge might interact differently with the tutor, for example, by asking more complex questions or relying less on guided hints.
 Furthermore, the study relied on self-reported survey data, which is subjective and may be influenced by their motivation and course experience.
Regarding external validity, the evaluation was limited to a four-week period and covered only a subset of the course topics (collections and functions). While this allowed for a controlled analysis of tutor usage in relevant problem areas, interaction patterns and perceived usefulness may differ for other topics, such as basic syntax or more advanced concepts. However, based on several years of teaching experience in the Python course, these topics were selected as those where students were expected to benefit most.
These limitations suggest that further studies are needed to evaluate the tutor across diverse learner populations, a broader range of topics, and extended course durations.

\section{Lessons Learned}
\label{sec:discussion}

Our experience with deploying and evaluating a course-specific, LLM-based tutor revealed several practical insights.

First, \textbf{context awareness proved to be a key factor for perceived usefulness}. The students especially liked that the tutor was able to consider both the task description, and their current code state and that \textbf{the explanations were grounded in the course materials}. This increased trust in the responses and further suggests that course-specific, context-aware tutoring provides clear advantages over generic LLM-based tools, particularly for challenging topics such as collections and functions.

Second, \textbf{guided, hint-based responses were generally well received}, even though students frequently asked for direct implementations. While interaction logs show quite a few solution-oriented requests, survey results indicate that students still perceived the tutor as supporting independent learning and active thinking. This suggests that a tutor can successfully focus on guidance rather than providing a complete solution, even when students initially try to get answers.

At the same time, \textbf{system responsiveness emerged as a limiting factor}. Multiple students mentioned that response latency somewhat interrupted their learning flow, despite otherwise positive evaluations. This highlights that performance characteristics, such as perceived responsiveness, can substantially influence user experience and should be considered a core design requirement, not only a technical detail. However, a fully scalable solution was out of scope for the prototype implementation.

Finally, students in our study often submitted \textbf{short or underspecified prompts}, including code-only inputs, implicitly expecting the tutor to infer intent from context. This highlights the importance of context handling and suggests that students quickly adapt their interaction behavior to conversational tutoring systems.

 \section{Conclusion and Future Work}
\label{sec:conclusion}

In this experience report, we have shown that course-aware AI tutoring can  support novice programmers by providing contextualized, hint-based guidance based on course materials. 
Students in our study reported that the tutor was helpful, especially due to its code- and task-aware responses and its step-by-step guidance. Analysis of student queries has also confirmed that while some students still tried to generate full solutions, the tutor was used for guided and individual problem solving and code comprehension. 
%As part of our future work, we will further explore how to improve usability, enhance system performance, and expand the capabilities of the tutor to a wider range of topics of our course.
Several directions for future work emerge from the limitations of the current study. First, the evaluation relied primarily on self-reported student perceptions collected through surveys. While these results indicate that students perceived the tutor as helpful, future work should investigate whether the system leads to measurable improvements in learning outcomes.
Second, the deployment covered only a four-week period and focused on a limited subset of course topics (collections and functions). Future work will therefore extend the tutor to additional topics across the entire course and investigate how interaction patterns and perceived usefulness evolve over a longer period of use.
Third, our study revealed that some students attempted to obtain complete solutions despite the tutor's goal of providing guided hints. A promising direction is to explore additional strategies, such as adaptive hint levels or mechanisms that detect repeated requests for solutions and instead encourage incremental problem solving.
Finally, further work is needed to improve the usability and responsiveness of the system.

\balance

\bibliographystyle{abbrv}
\bibliography{refs}

\end{document}